\documentclass[twocolumn,prb,floatfix,superscriptaddress,longbibliography]{revtex4-1}
\usepackage{amssymb}

\usepackage{graphicx}
\usepackage{dcolumn}
\usepackage{bm}
\usepackage{xspace}
\usepackage{hyperref}
\usepackage{color}
\usepackage[flushleft]{threeparttable}
\usepackage{multirow}

\def\new{\color{black}}
\def\fts{Fe$_{0.98}$Te$_{0.5}$Se$_{0.5}$\xspace}
\def\fcts{Fe$_{0.88}$Cu$_{0.1}$Te$_{0.5}$Se$_{0.5}$\xspace}

\begin{document}
\title{Enhanced low-energy magnetic excitations evidencing the Cu-induced localization  in an Fe-based superconductor Fe$_{0.98}$Te$_{0.5}$Se$_{0.5}$}
\author{Jinghui~Wang}
\affiliation{ShanghaiTech Laboratory for Topological Physics \& School of Physical
Science and Technology, ShanghaiTech University, Shanghai 200031, China}
\affiliation{National Laboratory of Solid State Microstructures and
Department of Physics, Nanjing University, Nanjing 210093, China}
\author{Song~Bao}
\author{Yanyan~Shangguan}
\author{Zhengwei~Cai}
\author{Yuan~Gan}
\author{Shichao~Li}
\affiliation{National Laboratory of Solid State Microstructures and
Department of Physics, Nanjing University, Nanjing 210093, China}
\author{Kejing~Ran}
\affiliation{ShanghaiTech Laboratory for Topological Physics \& School of Physical
Science and Technology, ShanghaiTech University, Shanghai 200031, China}
\affiliation{National Laboratory of Solid State Microstructures and
Department of Physics, Nanjing University, Nanjing 210093, China}
\author{Zhen~Ma}
\affiliation{Institute for Advanced Materials, Hubei Normal University, Huangshi 435002, China}
\author{B.~L.~Winn}
\affiliation{Neutron Scattering Division, Oak Ridge National Laboratory (ORNL), Oak Ridge, Tennessee 37831, USA.}
\author{A.~D.~Christianson}
\affiliation{Neutron Scattering Division, Oak Ridge National Laboratory (ORNL), Oak Ridge, Tennessee 37831, USA.}
\affiliation{Materials Science and Technology Division, Oak Ridge National Laboratory, Oak Ridge, Tennessee 37831, USA}
\author{Ruidan~Zhong}
\affiliation{Tsung-Dao Lee Institute, Shanghai Jiao Tong University, Shanghai 200240, China}
\affiliation{School of Physics and Astronomy, Shanghai Jiao Tong University, Shanghai 200240, China}
\author{Jun~Li}
\affiliation{ShanghaiTech Laboratory for Topological Physics \& School of Physical
Science and Technology, ShanghaiTech University, Shanghai 200031, China}
\author{Genda~Gu}
\affiliation{Condensed Matter Physics and Materials Science Department, Brookhaven National Laboratory, Upton, New York 11973, USA}
\author{Jinsheng~Wen}
\email{jwen@nju.edu.cn}
\affiliation{National Laboratory of Solid State Microstructures and Department of Physics, Nanjing University, Nanjing 210093, China}
\affiliation{Collaborative Innovation Center of Advanced Microstructures,
Nanjing University, Nanjing 210093, China}
\begin{abstract}
We have performed inelastic neutron scattering measurements on optimally-doped Fe$_{0.98}$Te$_{0.5}$Se$_{0.5}$ and 10\% Cu-doped Fe$_{0.88}$Cu$_{0.1}$Te$_{0.5}$Se$_{0.5}$ to investigate the substitution effects on the spin excitations in the whole energy range up to 300~meV. It is found that substitution of Cu for Fe enhances the low-energy spin excitations ($\le$ 100~meV), especially around the (0.5,\,0.5) point, and leaves the high-energy magnetic excitations intact. In contrast to the expectation that Cu with spin 1/2 will dilute the magnetic moments contributed by Fe with a larger spin, we find that the 10\% Cu doping enlarges the effective fluctuating moment from 2.85 to 3.13~$\mu_{\rm B}$/Fe, {\new although there is no long- or short-range magnetic order around (0.5,\,0.5) and (0.5,\,0)}. The presence of enhanced magnetic excitations in the 10\% Cu doped sample which is in the insulating state indicates that the magnetic excitations must have some contributions from the local moments, reflecting the dual nature of the magnetism in iron-based superconductors. We attribute the substitution effects to the localization of the itinerant electrons induced by Cu dopants. These results also indicate that the Cu doping does not act as electron donor as in a rigid-band shift model, but more as scattering centers that localize the system.
\end{abstract}
\maketitle

\section{INTRODUCTION}

In both copper- and iron-based high-temperature superconductors, understanding the interplay between superconductivity and magnetism has been a central issue\cite{kastner1998magnetic,tranquada1995evidence,tranquada2004quantum,dai2012magnetism,
dai2015antiferromagnetic,TRANQUADA2014148,si2016high}. In this regard, doping has acted as an extremely powerful tuning parameter. For example, for iron pnictides, rich phase diagrams have been obtained by substituting Fe with 3$d$ transition metals (TMs). With the isovalent doping of TM (TM=Co, Ni), the antiferromagnetic order in the parent compound is suppressed, and superconductivity appears with the superconducting temperature ($T_{\rm c}$)
$vs.$ doping having a dome shape\cite{PhysRevLett.104.057006,PhysRevLett.110.257001}. In the superconducting phase, a resonance peak in the paramagnetic excitations with the energy below twice of the superconducting gap is typically observed\cite{christianson2008unconventional,PhysRevB.82.172508,PhysRevB.79.174527,inosov2010normal}. An initial picture to understand the doping effect was the rigid-band model, which considered the extra $d$ electrons in the dopants contributing to the conduction bands\cite{liu2011importance,canfield2009decoupling}, and resulted in a rigid-band shift of the Fermi level\cite{neupane2011electron,nakamura2011first}. However, such a description has faced challenges from both theory\cite{wadati2010extra,berlijn2012transition} and experiment\cite{bittar2011co,kim2012effects,PhysRevB.85.214509,PhysRevB.90.024512}, as it ignores the impurity scattering induced by the dopants. For example, although Cu is next to Ni and Co, its doping effect is rather distinct from that of Co and Ni. With  increasing Cu concentration, samples can be driven into an insulating phase accompanied by the development of spin glass and long-range magnetic order\cite{PhysRevB.101.064410,song2016mott,PhysRevB.96.161106,PhysRevB.87.075105,PhysRevB.99.155114,PhysRevB.103.075112}. Remarkably, in a scanning tunneling microscopy study on NaFe$_{1-x}$Cu$_x$As, it is found that the local electronic structure of the insulating sample are strikingly similar to the Mott insulating phase of a lightly doped cuprate\cite{PhysRevX.5.021013,PhysRevB.99.155114}.

In another widely investigated iron-based superconductor system iron chalcogenide Fe$_{1+y}$Te, where superconductivity can be induced by substituting Te with Se~(Refs.~\onlinecite{katayama2010investigation,tranquada2020magnetism}), similar Cu substitution effects have been found. At both ends of the phase diagram of the Fe$_{1+y}$Te$_{1-x}$Se$_{x}$ system, the increasing doping of Cu will gradually drive Fe$_{1+y-z}$Cu$_{x}$Te~(Refs.~\onlinecite{wen2012magnetic,wang2012evolution}) and Fe$_{1+y-z}$Cu$_{x}$Se~(Refs.~\onlinecite{huang2010doping,williams2009metal}) from a metallic to an insulating phase, as well as induce a spin-glass state. Different substitution effects of Cu and Co/Ni on optimally-doped superconducting Fe$_{0.98}$Te$_{0.5}$Se$_{0.5}$ were also observed\cite{PhysRevB.91.014501}. Comparing with the impacts of Co or Ni substitution, the suppression of superconductivity and conductivity of Cu doping is more significant. Furthermore, with 10\% Cu doping, the resistivity shows a Mott-insulator behavior, which has been attributed to the stronger impurity potentials of Cu~(Ref.~\onlinecite{PhysRevB.88.144509}). Cu doping also enhances the low-energy~($\leq$12 meV) spin excitations significantly, {\new but without inducing a static magnetic order, either in the long-range or short-range form\cite{PhysRevB.88.144509,PhysRevB.91.014501}}. However, since these measurements were only performed at low energies, it is not clear that the spectral weight enhancement at low energies reflects the total enhancement or just a redistribution from high energies. In the former case, it means that the total fluctuating magnetic moment increases. While in the latter case, the fluctuating magnetic moments only slow down, and the total moment may remain unchanged or even decreased, consistent with the expectation for the diluting effect of Cu with a smaller spin. Therefore, inelastic neutron scattering (INS) measurements over the entire energy range will be required to make a definite conclusion.

In this paper, we investigate the effects of copper substitution on the spin excitations by performing comparative INS measurements on 10\% Cu doped \fcts (labeled as Cu10) and copper-free \fts (labeled as Cu0). With the large energy and momentum coverage of the time-of-flight spectrometers, we are able to uncover the full magnetic excitation spectrum. To analyze the data quantitatively, we have performed cross normalization of the data and obtained absolute values for the scattering intensities. From the normalized data, we find that the excitation up to $\sim$100~meV has been enhanced in Cu10, while the high-energy spectrum shows negligible difference. This gives rise to an effective moment of 3.13~$\mu_{\rm B}$/Fe in Cu10 than that of 2.85~$\mu_{\rm B}$/Fe in Cu0. Since Cu10 is insulating, the magnetic excitations must have contributions from local moments. On the other hand, the itinerant electrons are believed to give rise to incommensurate excitations including the magnetic resonance feature around (0.5,\,0.5) in Cu0~(Refs.~\onlinecite{qiu:067008,liupi0topp,PhysRevB.81.220503,hirschfeld2011gap,mazin2010superconductivity}). These results indicate the dual nature of the magnetic excitations and support that the Cu doping induces localization of the itinerant electrons and enhances the magnetic correlations.

\begin{table*}[tb]
\centering
\caption{\label{table1} Parameters of normalization for \fts (Cu0) and \fcts (Cu10) with the phonon data obtained on the HYSPEC spectrometer.}
\begin{tabular*}{\textwidth}{@{\extracolsep{\fill}} c c c c c c c c }
\hline
\hline
Sample & $\bm{Q}$ (rlu) & $E$~(meV) & $n_{\bm{q}}$ ($T=100$~K) & $\frac{\hbar\bf{Q}^2}{2m}$~(meV) & $\int \widetilde{I}(\bm{Q},E)dE$~(meV$^{-1}$) &   $|F_{\rm N}(\bm{G})|^2$ (b) & $\frac{1}{N\widetilde{R}_0}$ (meV$^{-1}$~b)\\
\hline
\multirow{2}{*}{Cu0} & (0,-2.140,0) & 5.000 & 2.258 & 25.94 & 0.0051  & 11.29 & 81.49 \\
 & (0,-1.830,0) & 4.928 & 2.280 & 18.97 & 0.0047  & 11.29 & 66.12 \\
\hline
\multirow{2}{*}{Cu10} & (0,-2.140,0) & 5.065 & 2.236 & 25.94 & 0.0041  & 11.00 & 95.34 \\
& (0,-1.860,0) & 4.410 & 2.480 & 19.60 & 0.0041  & 11.00 & 91.51\\
\hline
\hline
\end{tabular*}
\end{table*}

\section{EXPERIMENTAL DETAILS}
Single-crystal samples of \fcts and \fts (labeled as Cu10 and Cu0) were grown by the horizontal Bridgman method, as mentioned in our previous works\cite{PhysRevB.88.144509,PhysRevB.91.014501}. Each of the single crystals has a shape of semicylinder with two flat cleavage surfaces and a mass about 10~g. From resistivity measurement results\cite{PhysRevB.88.144509}, the Cu0 sample shows a $T_{\rm{c}}$ of 15~K, while the Cu10 sample is an insulator and the resistivity-temperature curve can be fitted well with a three-dimensional Mott variable range hopping formula.

The INS experiments were performed on the time-of-flight spectrometers ARCS and HYSPEC, both located at SNS of Oak Ridge National Laboratory. On ARCS, multiple incident energies of $E_{\rm i}=60, 180, 400$~meV were used with corresponding chopper frequency of 420, 600, and 420~Hz, respectively. The energy resolution for each $E_{\rm i}$ is about 5\% at $E=0$~meV. Since the wave vector $\bm{Q}$ and energy $E$ are coupled with fixed incident neutron momentum and sample orientation in the time-of-flight experiments, to cover a large range in the $(\bm{Q},E)$ space, the samples were rotated about the [010] axis by 90$^{\circ}$ with a step of 5$^{\circ}$ on ARCS. For the \fts sample, extra data with a step of 1$^{\circ}$ for $E_{\rm i}=60$ and 180~meV were also collected. Data were further folded and averaged along the [100] axis to reduce noise and instrument-induced streaks by DAVE software. On HYSPEC, an incident energy of $E_{\rm i}=35$~meV and a Fermi chopper frequency of 240~Hz was used with the energy resolution $\Delta E \approx 3$~meV at $E=0$~meV. Both samples were rotated in the $a$-$b$ plane by 90$^{\circ}$ with a step of 1$^{\circ}$ on HYSPEC. The data obtained on HYSPEC covered approximately one quadrant of the ($H,\,K,\,0$) plane and had been symmetrized to be four-fold to be compared with the higher-energy data collected on ARCS. On both spectrometers, the crystals were mounted on aluminum sample holders and loaded into a closed-cycle refrigerator. All neutron scattering measurements were performed at 100~K, which was well above the $T_{\rm{c}}$ of \fts to avoid any effects from the superconducting correlations. When presenting the INS data in the reciprocal space, we used the configuration of two-Fe unit cell, of which the lattice constants at room temperature were $a=b\approx3.8$~{\AA} and $c=6.1$~{\AA} for both samples. The wave vector $\bm{Q}$ was expressed as ($H,\,K,\,L$) reciprocal lattice unit (rlu) of $(a^{*},\,b^{*},\,c^{*})=(2\pi/a,\,2\pi/b,\,2\pi/c)$.

As the magnetic excitations in Fe$_{1+y}$Te$_{1-x}$Se$_{x}$ are reported to be of two-dimensional nature because of the much weaker interplanar correlations\cite{lumsden2010evolution}, we had projected the spin excitations onto the ($H,\,K,\,0$) plane with $L=[-2,2]$ to improve the statistics for the data obtained from ARCS. For the $E_{\rm i}=180$ and 400~meV data, the combining range was $L=[-4,4]$. To correctly project the large-$L$ intensity into the ($H,\,K,\,0$) plane, we also applied a correction to each data point with the $\bm{Q}$-dependent isotropic Fe$^{2+}$ magnetic form factor. For the ARCS data, intensity from $(0,~0,~L)$ was used as the background, while for the HYSPEC data, background intensity was obtained at (0.5,\,0,\,0) for subtraction.

\begin{figure*}[htb]
\centering
\includegraphics[width=0.9\linewidth]{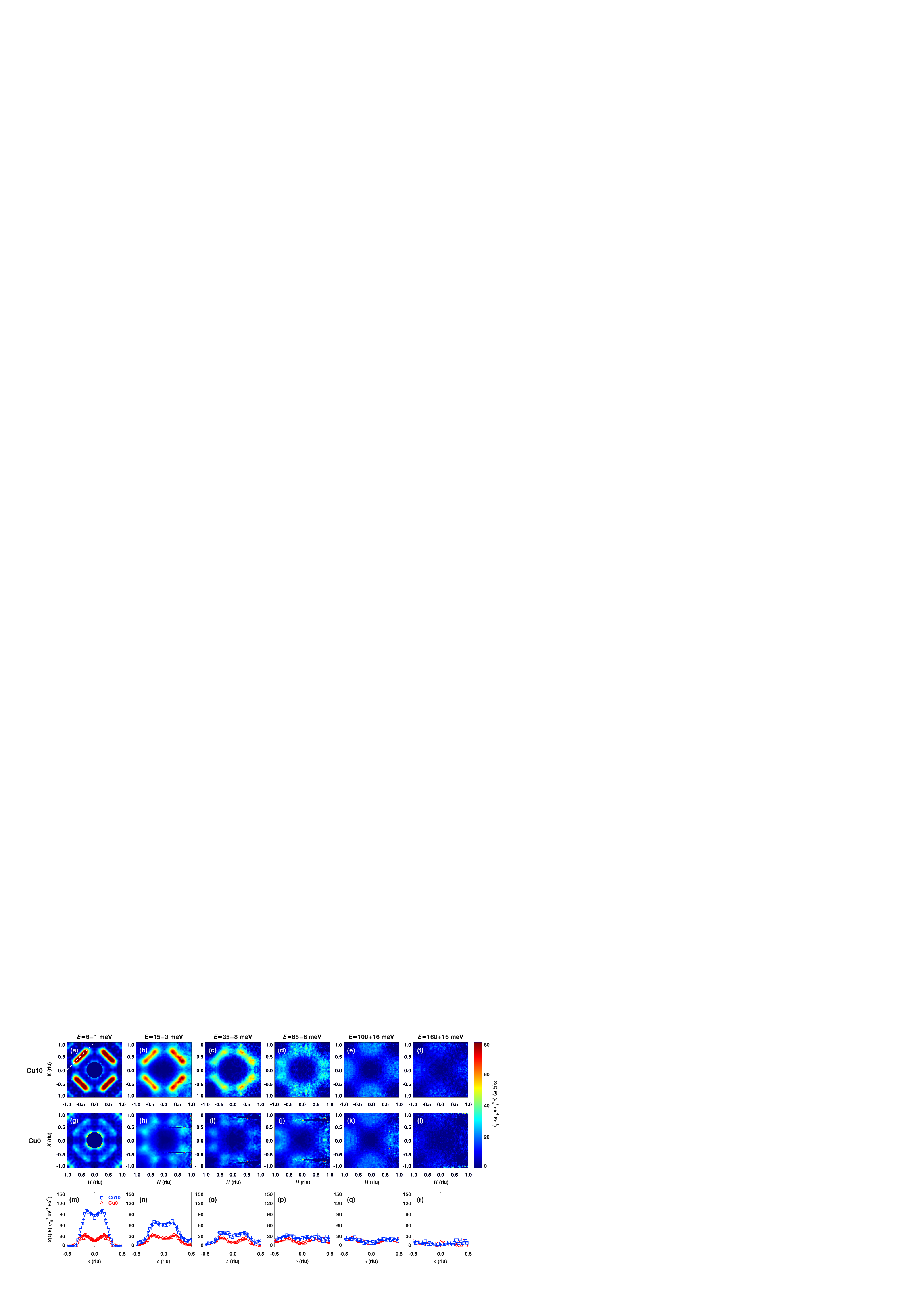}
\caption{{\textbf{Constant-energy contour maps and line cuts of the magnetic excitations at 100~K for \fcts and \fts.} Panels \textbf{(a-f)} and \textbf{(g-l)} present constant-energy contours for Cu10 and Cu0 samples, respectively. The measurements in \textbf{(a)} and \textbf{(g)} were carried out on HYSPEC and the data were symmetrized to be four-fold symmetric, while the data on ARCS in the other panels were folded along the $H$ axis to the $K\ge0$ side and then the folded data were duplicated on the $K\le0$ side to represent the four-fold symmetry. The streaks across the excitations in panels \textbf{(g)}, and \textbf{(h)} to \textbf{(j)} were due to the lack of detector coverage wherein. Panels \textbf{(m-r)} are linear cuts of the spin excitations through (-0.5,\,0.5) along the [110] direction, illustrated as the dashed line in panel {\bf (a)}, at the corresponding energies labeled in the same column. The errors through data points represent one standard deviation throughout the paper. Solid lines through data are fitted with Gaussian functions. The data integration range for the linear cuts are $H=K=\pm0.025$~rlu.}}
\label{fig:fig1}
\end{figure*}

A prerequisite process of quantitatively comparing the magnetic excitations of different samples is to normalize their intensities. In practice, the measured intensity should be converted to the dynamical spin-spin correlation function $S(\bm{Q},E)$ with the absolute unit of $\mu{_{\rm{B}}}^2\,{\rm eV}^{-1}$/Fe, which has been introduced in Ref.~\onlinecite{Guangyong2013Absolute} in detail. In this work, we measured the acoustic phonon branches to perform the cross normalizations. To numerically describe $S(\bm{Q},E)$ by the measured four-dimensional intensity $I(\bm{Q},E)$, one has:
\begin{equation}
S(\bm{Q},E)=\frac{13.77({\rm b}^{-1})I({\bm{Q}},E)}
{g^2|f(\bm{Q})|^2{\rm e}^{-2W}N\widetilde{R}_0},
\label{equation1}
\end{equation}
where 1~${\rm b}=10^{-24}$~cm$^2$ is the unit for neutron scattering crosssection. Here, $g$, $f({{\bf Q}})$, and ${\rm e}^{-2W}$ are Land$\acute{e}$ $g$-factor,  magnetic form factor, and Debye-Waller factor, respectively. The instrument resolution volume $N\widetilde{R}_0$ can be obtained by measuring acoustic phonons:
\begin{equation}
\frac{1}{N\widetilde{R}_0}=\frac{n_{\bm q}}{E({\bm q})}\frac{(\hbar\bm{Q})^2}{2m}\frac{m}{M} \frac{\cos^2\beta|F_{\rm N}(\bm{G})|^2{\rm e}^{-2W}}{\int \widetilde{I}(\bm{Q},E)dE}.
\label{equation2}
\end{equation}
Here ${\bm q}={\bm Q}-{\bm G}$ is the reduced wave vector, and $n_{\bm q}=1/(1-{\rm e}^{-E/k_{\rm B}T})$ is the Bose factor. $m$, $M$, $F_{\rm N}(\bm{G})$, and $\beta$ are the neutron mass, atomic mass of one unit cell, the acoustic phonon structure factor at Bragg peak $\bm{G}$ near $\bm{Q}$ at which the acoustic phonon is measured, and the angle between $\bm{Q}$ and phonon polarization direction. By performing constant-$\bm{Q}$ scans, one can fit the scans and obtain the final integrated phonon intensity $\int \widetilde{I}(\bm{Q},E)dE$. To improve the reliability of our data, we chose the data from HYSPEC with better energy resolution as the reference points for cross-normalization. The data on ARCS had also been normalized with phonons measured with $E_{\rm i}=60$~meV and then cross checked with the HYSPEC data where they had overlapping energy ranges. The determined parameters for the normalization of the HYSPEC data are presented in TABLE~\ref{table1}. Averaged values of the resolution volume $N\widetilde{R}_0$ at different $\bm{Q}$s were taken when calculating $S(\bm{Q},E)$.

\section{RESULTS AND ANALYSES}
\subsection{Spin excitations}

\begin{figure}[htb]
\centering
\includegraphics[width=\linewidth]{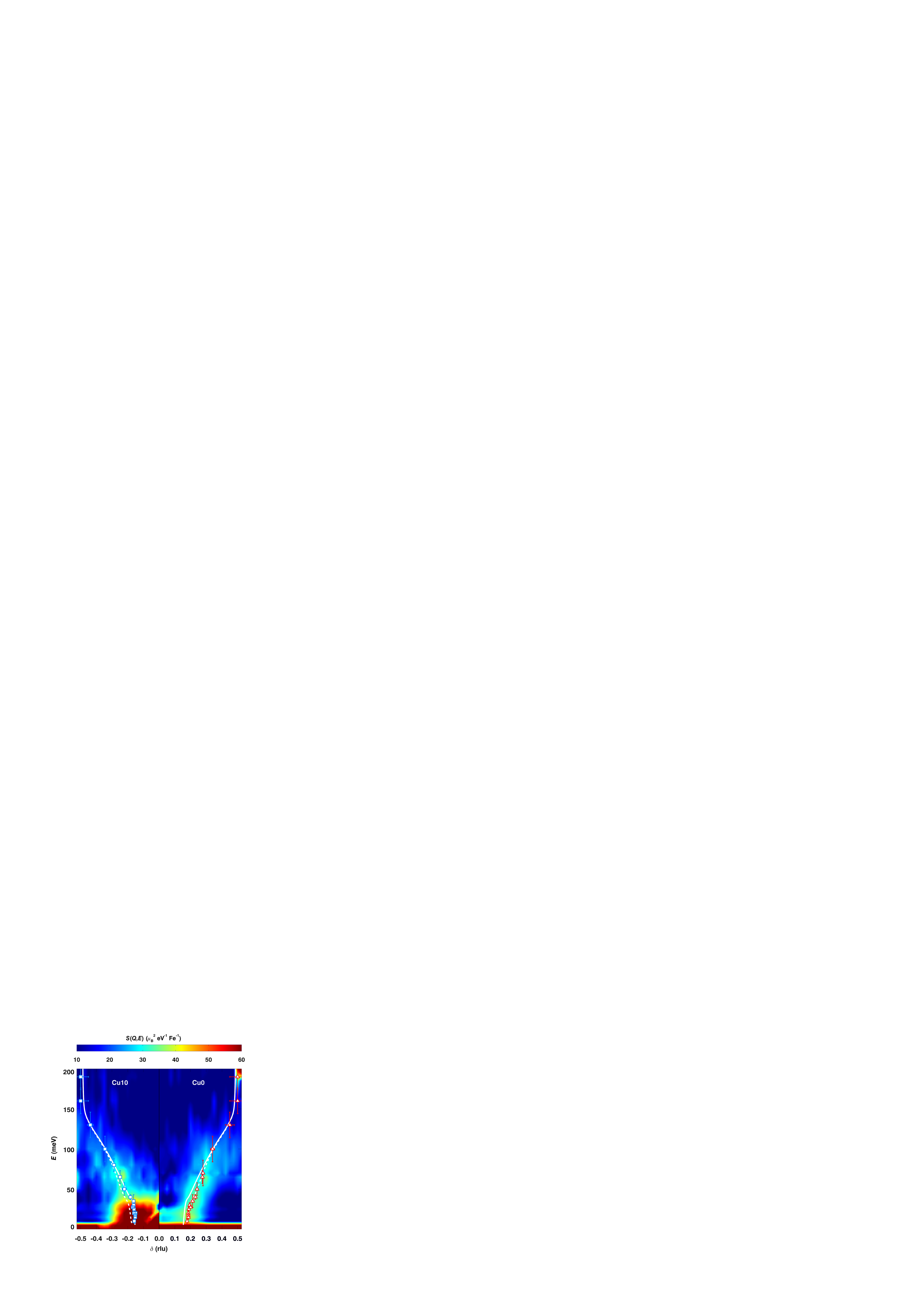}
\caption{{\textbf{Dispersions for \fcts (Cu10) and \fts (Cu0).} Left and right panels are contour maps of the dispersions presented by energy evolutions of the intensity along ($-0.5\pm\delta$,\,$0.5\pm\delta$) for Cu10 and Cu0 samples, respectively. The integration range along the [1$\bar{1}$0] direction is 0.056~rlu. For each sample, only the dispersion on one side of (-0.5,\,0.5) has been presented, with the other side folded and averaged for both panels. The data points on top of the contour maps are the peak positions extracted from the fittings with Gaussian functions to the constant-$E$ scans such as those shown in Fig.~\ref{fig:fig1}\textbf{(m)-(r)}. The vertical errorbars stand for the energy binning ranges used for making the fittings, and the horizontal bars are the fitting errors. Solid and dashed lines through data are guides to the eye, illustrating the dispersions for Cu10 and Cu0, respectively.}}
\label{fig:fig2}
\end{figure}

To understand how the spin excitations of Cu10 sample evolve with energy, we begin with a series of constant-energy cuts of the magnetic spectra as shown in Fig.~\ref{fig:fig1}(a)-(f) with  energy transfers up to 160~meV. Results for the Cu0 sample are arranged in the second row as Fig.~\ref{fig:fig1}(g)-(l), which are found to be consistent with previous measurements on a sample with similar composition FeTe$_{0.51}$Se$_{0.49}$, after considering the effect of magnetic form factor at large $\bm{Q}$s~(Ref.~\onlinecite{lumsden2010evolution}). The linear-cut comparisons of Cu10 and Cu0 through (0.5,\,-0.5) along the [110] direction are presented in the third row as Fig.~\ref{fig:fig1}(m)-(r).

From the energy slices of both samples, we can find differences in the peak intensities and pattern shapes at energies below $\sim$65~meV. At higher energies, the scattering appears to be quite similar. Specifically, taking the second-quadrant data as representatives, two widely studied incommensurate peaks located at ($-0.5\pm\delta$,\,$0.5\pm\delta$) elongating along the [110] direction can be recognized for both Cu10 and Cu0 samples at 35~meV or lower energies. Here $\delta$ represents the incommensurability displaced from (-0.5,\,0.5) along the [110] direction. The peak intensities of Cu10 sample are significantly larger than those of Cu0 sample, and there appears to be more spectral weight filling in around (-0.5,\,0.5) for Cu10, making the scattering more like a rod elongating along [110], instead of two well-separated incommensurate peaks as in Cu0. These results are consistent with our previous data at low energies\cite{PhysRevB.88.144509}. As the energy increases, the spectra for both samples become convergent: the incommensurate peaks begin to disperse further away from (-0.5,\,0.5) along the [110] direction, and gradually approach (-1,\,0) and (0,\,1). In the high-energy spectrum ($E\ge 100$~meV), the excitations centering (-1,\,0) and (0,\,1) become diffusive in $\bm{Q}$. The overall pattern does not change further at higher energy, albeit that the spectral weight gradually decreases with increasing energy and diminishes eventually.

To follow the change of the excitations in more detail, we perform $\bm Q$ scans in the second-quadrant of the constant-energy cuts in Fig.~\ref{fig:fig1}(a)-(l) along the [110] direction, and plot the results corresponding to each energy transfer in the third row as Fig.~\ref{fig:fig1}(m)-(r). The intensity enhancement below 100~meV in the Cu10 spectrum is again clearly demonstrated, extending the conclusion obtained from the triple-axis data to a much larger energy scale\cite{PhysRevB.88.144509}. For both samples, the excitations are clearly incommensurate about (-0.5,\,0.5). However, the incommensurability becomes smaller in Cu10, along with more spectral weight filling into  (-0.5,\,0.5), making the relative intensity difference at the incommensurate peaks and (-0.5,\,0.5) smaller. 

To further characterize the dispersion, we have obtained a series of such $\bm Q$ scans from 6 to 190~meV, and plot the dispersions along the [110] direction for both samples in Fig.~\ref{fig:fig2}. At each energy, we performed the same fitting as the lines shown in Figs.~\ref{fig:fig1}(m)-(r) and obtained the peak positions. As the two branches of the dispersions with positive and negative $\delta$ may not be equivalent, we have averaged $|\delta|$ for each sample in the figure. A notable separation between the dispersions of Cu10 and Cu0, with the peaks of Cu10 being closer to (0.5,\,0.5), is obvious below $\sim$100~meV. Above 100~meV, the dispersions for the two samples merge together. Above 160~meV, the excitations persist at around (1,\,0) and (0,\,1) without further dispersing outwards. The overall features are similar to those of FeTe$_{0.51}$Se$_{0.49}$~(Ref.~\onlinecite{lumsden2010evolution}), and remarkably, the high-energy stripe-like excitations are similar to those in the parent compound Fe$_{1+y}$Te~(Refs.\onlinecite{spinwave2011,stock2014soft}). However, before the incommensurate peaks gradually disperse to (1,\,0) and (0,\,1) at high energies, a kink occurs at about 40~meV for Cu10 and 30~meV for Cu0. Below the kink energy, the incommensurate peaks are almost dispersionless, and above it, the excitations become dispersive. We suspect the kink energy to be the characteristic energy scale that distinguishes two types of excitations---the paramagnetic excitations around (0.5,\,0.5) at low energies and the remnant of the spin waves in the parent compound Fe$_{1+y}$Te at high energies\cite{spinwave2011,stock2014soft}. Compared with the dispersion of the Cu0 sample, the dispersion Cu10 is steeper at low energies, and the kink energy is also higher. This indicates that the ($\pi,\,\pi$)-type spin correlation in the Cu10 sample is more robust, consistent with the spectral weight enhancement around (-0.5,\,0.5) as discussed above.

\subsection{Magnetic moments}

\begin{figure}[htb]
\centering
\includegraphics[width=\linewidth]{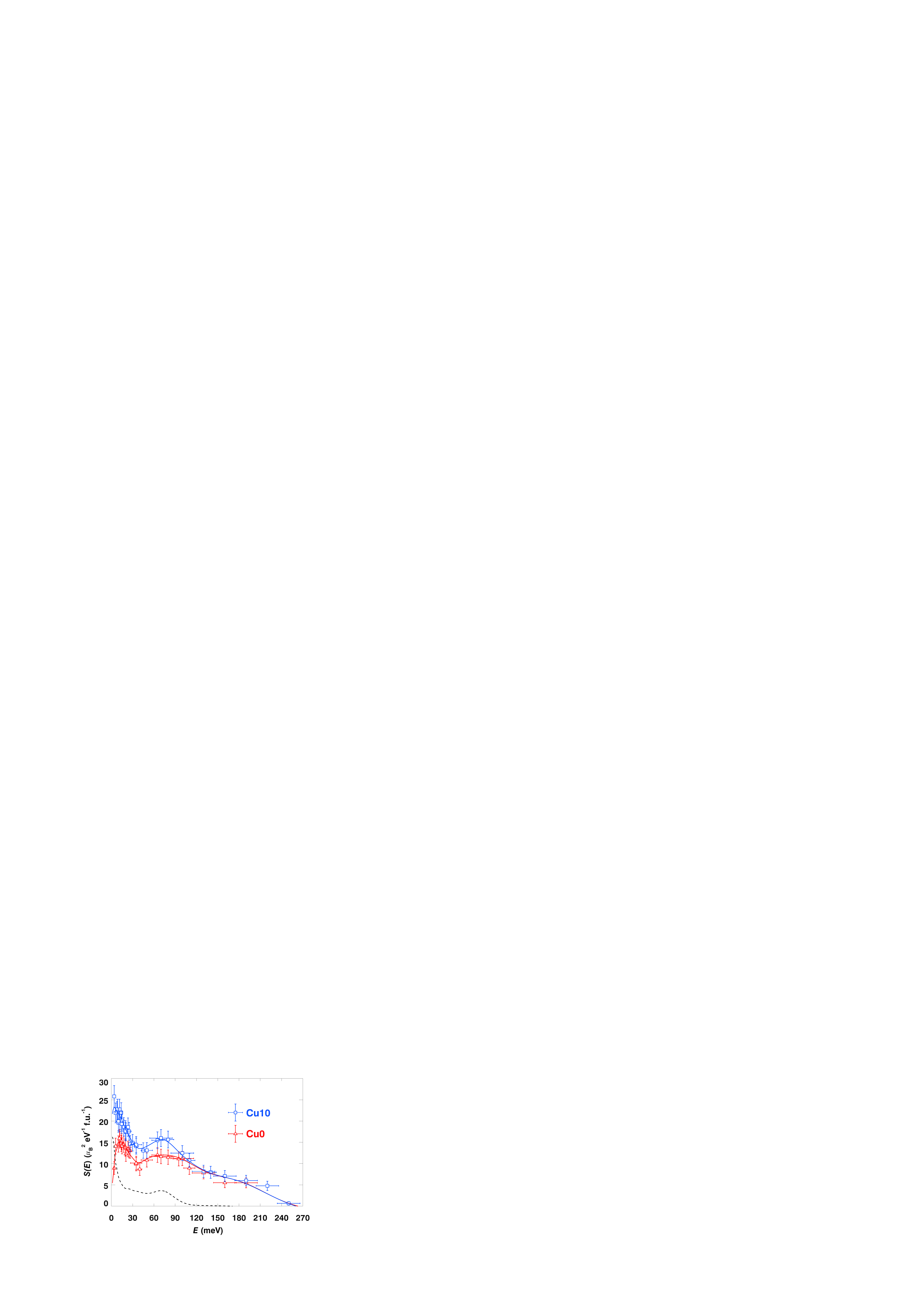}
\caption{{\textbf{Energy dependence of the $\bm{Q}-$integrated dynamical spin-spin correlation function $S(E)$ for \fcts and \fts.} Integration of $S(E)$ covers an area of one Brillouin zone (-1$\le$H$\le$0,\,0$\le$K$\le$1). The normalized data from different spectrometers and incident energies are consistent with each other in the overlapping regimes, indicating the reliability of the normalizations. Horizontal errorbar stand for the energy binning range, and vertical error represents one standard deviation. Solid lines through data are guides to eye. The dashed line is plotted by subtracting intensities of Cu0 from Cu10, indicating the extra intensities induced by the Cu substitution.}}
\label{fig:fig3}
\end{figure}

To further investigate the evolution of the magnetic excitations with doping, we have calculated the wave-vector-integrated correlation function $S(E)$ by integrating the intensity in one Brillouin zone. Since the data are nosier in the $H\ge0$ zone, we choose the second quadrant (-1$\le$H$\le$0, 0$\le$K$\le$1) as the integrated area. In Fig.~\ref{fig:fig3}, we show the evolution of $S(E)$ with energy, where the data obtained with different incident energies and from different spectrometers have been plotted together. At the overlapped energies, the data from different energies coincide with each other, indicating the high reliability of the absolute cross-section normalizations.

In addition to the enhancement of the peak intensity from the linear cuts, the $\bm Q$-integrated correlation function $S(E)$ also exhibits an enhancement with the Cu doping. The enhancement induced by the Cu substitution is sketched by the dashed guide line. The existence of a cut-off energy $\sim$100~meV above which $S(E)$ is the same for both samples is in contrast to the idea that the enhancement happens homogeneously in the whole energy range. This cut-off energy is in accordance with the energy where the dispersions of Cu10 and Cu0 samples overlap as shown in Fig.~\ref{fig:fig2}. It is also worth mentioning that the high-energy spectrum is similar to that of the parent compound Fe$_{1+y}$Te~(Refs.~\onlinecite{zaliznyak2011unconventional,stock2014soft}), which implies that the remnant of the spin waves remains unchanged in both samples. Such phenomenon is similar to the case in the electron-doped BaFe$_2$As$_2$~(Refs.~\onlinecite{wang2013doping,liu2012nature}), where the high-energy spectrum remains unchanged while only the low-energy excitation (below $\sim$100~meV) is influenced.

With the integrated correlation function $S(E)$, we can also calculate the fluctuating instantaneous effective moment $\mu_{\rm{eff}}$ immediately, by the sum rule and their relationship\cite{Guangyong2013Absolute,zaliznyak2011unconventional}:
\begin{equation}
\mu_{\rm{eff}}^2=g^2\mu{_{\rm{B}}^2}\int S(\bm{Q},E)d\bm{Q}dE=g^2\mu{_{\rm{B}}^2}S(S+1),
\label{equation3}
\end{equation}
where Land$\acute{e}$ factor $g=2$ and $S$ is the effective local spin. The determined $\mu_{\rm{eff}}$ for Cu10 and Cu0 samples at 100~K are $3.13\pm0.06$~$\mu{_{\rm{B}}}$/Fe
and $2.85\pm0.06$~$\mu{_{\rm{B}}}$/Fe, respectively. For comparison, the effective moment of both samples are smaller than $\mu_{\rm{eff}}\sim3.7~\mu{_{\rm{B}}}$/Fe in the parent compound Fe$_{1+y}$Te from Refs.~\onlinecite{zaliznyak2011unconventional,stock2014soft}, but larger than the $\sim$1.1~$\mu{_{\rm{B}}}$/Fe for FeSe~(Ref.~\onlinecite{wang2016magnetic}) at 110~K.
The effective local spin $S$ is $1.14\pm0.02$ for Cu10 and $1.01\pm0.02$ for Cu0, which is close to an $S=1$ ground state with the existence of itinerant electrons at 100~K.

\section{DISCUSSIONS}

From our comprehensive magnetic excitation spectra in the whole energy scale on both Cu10 and Cu0 samples, we have found that while the high-energy magnetic excitations, which are likely to be the remnant of the spin waves from the parent compound, remain intact, the low-energy excitations ($<100$~meV) are substantially enhanced. In addition to the intensity enhancement at the incommensurate peaks in the Cu10 sample at low energies, there is more spectral weight filling in around the commensurate position (0.5,\,0.5). The velocity of the low-energy excitations and the kink energy where the low-energy excitations become dispersive are both larger in the Cu10 sample, indicating the strengthened magnetic interactions. As a result, the effective magnetic moment increases from $2.85\pm0.06$~$\mu{_{\rm{B}}}$/Fe in Cu0 to $3.13\pm0.06$~$\mu{_{\rm{B}}}$/Fe in Cu10, corresponding to an effective spin of $1.01\pm0.02$ and $1.14\pm0.02$ for Cu0 and Cu10, respectively. These results are in contrast to the expectation that the Cu doping will dilute the magnetic moment. Instead, it is indicated that the Cu doping will enhance the magnetic correlations around (0.5,\,0.5). This resolves the uncertainties of the low-energy data in our previous works\cite{PhysRevB.88.144509,PhysRevB.91.014501}.

We believe the enhancement of the low-energy magnetic excitations around (0.5,\,0.5) is at the expense of the itinerancy of the Fe electrons. In our previous works, we have shown from the resistivity measurements that the Cu doping will suppress the itinerancy of the system dramatically\cite{PhysRevB.88.144509,PhysRevB.91.014501}---with a 10\% Cu doping, the Cu10 sample becomes a Mott insulator effectively. These localized electrons partly contribute to the fluctuating magnetic moments, resulting in an enhancement in the effective magnetic moment as well as the local spin. Since the magnetic excitations around (0.5,\,0.5) are not only present but also enhanced when the system is already in the insulating state, these excitations must not be purely resulting from the Fermi surface nesting as in a weak-coupling picture, otherwise they should be diminishing in Cu10. This is in line with our previous work which showed that the Ni doping suppressed the itinerancy but not the magnetic excitations in Fe$_{0.98-z}$Ni$_z$Te$_{0.5}$Se$_{0.5}$~(Ref.~\onlinecite{PhysRevB.91.014501}). These results reflect the dual nature of the magnetic excitations in these systems. Due to the different characters of the Fe $d$ orbitals\cite{Cvetkovic_2009,PhysRevB.90.165123,dai2012magnetism,PhysRevLett.101.126401}, some localized electrons contribute to the local moments, while the itinerant electrons on the Fermi surface may also contribute to
paramagnetic excitations around (0.5,\,0.5) by the nesting between the hole pockets at the $\mit\Gamma$ point and the electron pockets at the $M$ point---such a picture appears to well explain the neutron-spin resonance as a quasi-particle exciton\cite{PhysRevB.81.220503,hirschfeld2011gap,mazin2010superconductivity}. In fact, the presence of different characteristic energies in the dispersions shown in Fig.~\ref{fig:fig2} is a vivid demonstration that there are several different and possibly competing magnetic interactions in the systems\cite{stock2014soft}. In this work, it is shown that Cu doping acts as an effective tuning parameter of the competing interactions. {\new Along this line, we suspect that if samples with higher Cu dopings become available, a spin-glass or magnetically ordered ground state, which is absent in Cu10, may be achieved\cite{PhysRevB.101.064410,song2016mott,PhysRevB.96.161106,PhysRevB.87.075105,PhysRevB.99.155114,PhysRevB.103.075112,huang2010doping,williams2009metal}. }These results further suggest that in describing the magnetism of the transition-metal compounds, while both the itinerant-electron and local-moment picture have their own merits or drawbacks, a more appropriate approach seems to be considering the contributions from both components as well as the interactions between them\cite{PhysRevX.12.011022,kim2022kondo}.

Overall, our results indicate that the rigid-band shift model is not universally applicable in describing the substitution effect. One of the ingredients that needs to be taken into account is the different scattering potentials of the dopants. For example, from Co to Ni and to Cu, the scattering potential of the element is increasing. As a result, the suppression on the superconductivity and enhancement on the low-energy magnetic excitations become more significant. Another factor that makes the doping effect more complicated is the presence of multiple Fe $d$ orbitals. Because of their different characters, how they respond to the doping can also be different. {\new Furthermore, Co, Ni, and Cu have roughly the same size, but Cu is Jahn Teller active and possibly can induce a different local distortion in the host lattice compared to Co and Ni~(Ref.~\onlinecite{RevModPhys.60.585}). Therefore, the Jahn Teller distortion can also play some role, which is to be examined with more detailed structural studies.}

\section{Conclusions}
In summary, we have investigated the substitution effects of transition metal Cu in the iron-based superconductor \fts. It is found that, with 10\% Fe substituted by Cu atoms, the low-energy magnetic excitations up to $\sim$100 meV are significantly enhanced, while the high-energy spectra show negligible difference. The Cu substitution induces an enhancement of the effective moment from $2.85\pm0.06$~$\mu{_{\rm{B}}}$/Fe in Cu0 to $3.13\pm0.06$~$\mu{_{\rm{B}}}$/Fe in Cu10, which correspond to a spin of of $1.01\pm0.02$ and $1.14\pm0.02$, respectively. The enhancement is at the cost of the itinerancy of the Fe electrons. These results depict the dual nature of magnetic excitations and interesting and complex doping effect beyond the rigid-band shift model in Fe$_{1+y-z}$Te$_{1-x}$Se$_x$ in specific, and likely in iron-based superconductors in general.

\section{ACKNOWLEDGEMENTS}

The work was supported by National Key Projects for Research and Development of China with Grant No.~2021YFA1400400, the National Natural Science Foundation of China with Grant Nos. 12074174, 12004251, and 12004249, Hubei Provincial Natural Science Foundation of China with Grant No.~2021CFB238, Shanghai Sailing Program with Grant Nos.~21YF1429200  and 20YF1430600, Fundamental Research Funds for the Central Universities. The work at Brookhaven National Laboratory was supported by the US Department of Energy, Office of Basic Energy Sciences under Contract No.~DOE-SC0012704. A portion of this research used resources at the Spallation Neutron Source, a DOE Office of Science User Facility operated by the Oak Ridge National Laboratory. ADC was partially supported by the U.S. Department of Energy, Office of Science, Basic Energy Sciences, Materials Science and Engineering Division. Z.M. thanks Beijing National Laboratory for Condensed Matter Physics for funding support.

%



\end{document}